\documentclass[prl,twocolumn,showpacs,floatfix,amsmath,amsfonts]{revtex4}

\usepackage{dcolumn}

\begin{document}
\title{Landau dynamics of a grey soliton in a trapped condensate}
\author{Vladimir V. Konotop$^{1}$ and Lev Pitaevskii$^{2}$
}
\address{$^{1}$Centro de F\'{\i}sica Te\'orica e Computacional,
Universidade de Lisboa, Complexo Interdisciplinar, Avenida
Professor Gama Pinto 2, Lisboa 1649-003, Portugal.
\\
$^{(2)}$Dipartimento di Fisica, Universit\`{a} di Trento and
Istituto Nazionale per la Fisica della Materia, CRS-BEC, 38050
Trento, Italy, and Kapitza Institute for Physical Problems, 119334
Moscow, Russia. }

\begin{abstract}
\par
It is shown that  grey soliton dynamics in  an one-dimensional
trap can be treated as Landau dynamics of a quasi-particle. A
soliton of arbitrary amplitude moves in the trapping potential
without deformation of its density profile as a particle of  mass
$2m$. The dynamics in the local density approximation is shown to
be consistent with the perturbation theory for dark solitons.
Dynamics of a vortex ring in a trap is discussed qualitatively.

\end{abstract}
 \maketitle



Following the experimental observation of grey solitons (or more
precisely, entities associated with quasi-one-dimensional grey
solitons) in Bose-Einstein condensates (BEC's)~\cite{experim}, a
great deal of attention has been paid to the theory of the
phenomenon (see e.g. \cite{Anglin,BK1} and references therein). It
was found that a grey soliton in a parabolic trap displays a
number of peculiarities. Among them, we mention the frequency of
oscillations, which is $\sqrt{2}$ times less than the frequency of
oscillations of the condensate as a whole~\cite{Anglin}, and the
consequent beatings, which can be observed in the long-time
dynamics of a soliton, an internal mode accompanying soliton
dynamics, nontrivial phase changes in soliton
evolution~\cite{BK1}, etc. It turns out, however, that the
mathematical treatment of the problem, based on the application of
perturbation theory to dark solitons~\cite{KV94} and covering all
the main effects, is rather involved, and more important, requires
a small   soliton velocity. The major problem is that a grey
soliton even in the simplest one-dimensional (1D) parabolic trap
is dramatically different from the standard grey soliton known
from soliton theory. Distinctions emerge from different boundary
conditions, which are zero in the case of a trap potential and
nonzero in the case of a mathematical grey soliton (see
e.g.~\cite{FT}).

The present paper aims to describe the motion of a grey soliton
over a wide range of velocities in a trapped condensate, whose
longitudinal size is sufficiantly large. The phenomenon  is
described by the  mean-field 1D Gross-Pitaevskii (GP) equation
\begin{eqnarray}
\label{GP}
i\hbar\Psi_t=-\frac{\hbar^2}{2m}\Psi_{xx}+U(x)\Psi+g_1|\Psi|^2\Psi
\;.
\end{eqnarray}
We consider a cigar-shaped condensate with  the following
relations among the transverse $a_\bot$, the longitudinal $2L$
dimensions of the condensate and a healing length $\xi
=\frac{\hbar }{\sqrt{2}mc}$ ($c$ hereafter being the sound
velocity): $a_\bot\ll \xi \ll 2L$. The 1D coupling constant $g_1$,
comes from the ``averaging'' over the transverse cross section of
the condensate~\cite{g1D,BK1}.

We developed a full description of the dynamics of the soliton,
considering it as a quasiparticle and using the Landau theory of
superfluidity. Subsequently we show that this approach is
consistent with the GP equation and derive the main result by use
of the proper perturbation procedure. In conclusion, on the basis
of the results obtained we discuss qualitatively the dynamics of a
vortex ring.

\paragraph{Soliton in a homogeneous condensate.}

We start with a gray soliton in a uniform 1D condensate, i.e. when $U(x)\equiv 0$.
According to Tsuzuki~\cite{Tsuzuki} the respective condensate wave function can
be presented in the form (see also~\cite{Pit1}, \S 5.5):
\begin{eqnarray}
\Psi \left( x,t \right ) =\sqrt{n}\left( i\frac{v}{c}+ \frac{u}{c}
\tanh \left[ \frac{1}{\ell}(x-X(t)) \right] \right) e^{-i\mu
t/\hbar}\;, \label{soliton}
\end{eqnarray}
where $X(t)=vt$, $v$ is the velocity of the soliton, $n$ is the
unperturbed linear density,  $\mu=g_1n=mc^2$ is the chemical
potential, $u=\sqrt{c^2-v^2}$, and $\ell=\hbar/mu$ is the width of
the soliton.
The energy of the soliton is computed to be
\begin{eqnarray}
{\mathcal E}  =\frac{4\hbar m}{3g_{1}}\left( c^{2}-v^{2}\right)
^{3/2}=\frac{4\hbar m}{3g_{1}}u^3\;. \label{Es}
\end{eqnarray}

\paragraph{Landau dynamics of a soliton in an inhomogeneous condensate.}

Let us now suppose that the condensate length, $2L$, is large
compared to the width of the soliton:
\begin{eqnarray}
L\gg \ell \;.
 \label{size}
\end{eqnarray}
Then, for such large condensates one can use semiclassical Landau
dynamics of the soliton, where the quantity $\left(
\ref{Es}\right) $ plays role of the Hamiltonian of quasiparticle.
We notice that condition (\ref{size}) is  stronger than the
condition of applicability of the {\it Thomas-Fermi} (TF)
approximation to the condensate $L\gg \xi $~\cite{Pit,Pit1}.
Requirement (\ref{size}) ensures, also, that the velocity of
motion of the condensate as a whole due to the soliton
oscillations is small and one can regard  the condensate as being
in the rest.

Subject to condition (\ref{size}), one can use a
{\em local density approximation}, assuming that equation
(\ref{Es}) for the soliton energy is valid in the inhomogeneous
condensate, i. e.,  that $c$ can be changed to its local value
$c\left( X\right)$, where $X$ is the position of the center of the
soliton. In the first approximation the soliton wave function has
the same form (\ref{soliton}),
where $X$ and $v$ are functions of time related by
${dX}/{dt}=v(t)$.
The soliton motion is then defined by the energy conservation
equation
\begin{eqnarray}
\frac{4\hbar m}{3g_{1}}\left[ c\left( X\right) ^{2}-v^{2}\right]
^{3/2}={\cal E} =const
\end{eqnarray}
which obviously implies that $u$ is also a constant.
Expressing ${\cal E}$ in terms of $u$,  we finally find  the
equation
\begin{eqnarray}
\left( \frac{dX}{dt}\right) ^{2}=c\left( X\right) ^{2}-u^2 \;
\label{dzdt}
\end{eqnarray}
which can be solved by a simple integration.

In order to find the distribution of the sound speed, we
approximate the density by the TF  law
\begin{eqnarray}
n_{TF}(x)=\left( \mu -U\left( x\right) \right)/g_{1}\;.\label{nTF}
\end{eqnarray}
Assuming, without loss of generality, that $U\left( 0\right) =0$,
we find
\begin{eqnarray}
c\left( X\right) ^{2}=g_1n_{TF}/m=c_{0}^{2}-U\left( X\right) /m
\end{eqnarray}
where $c_{0}=\sqrt{\mu /m}$ is the sound speed at $x=0$.
Substitution into $\left( \ref{dzdt}\right) $ gives
\begin{equation}
m\left( \frac{dX}{dt}\right) ^{2}+U\left( X\right)
=m(c_{0}^{2}-u^2)\;.\label{final}
\end{equation}
This equation describes classical motion of a particle having mass
$2m$ and energy $m(c_{0}^{2}-u^{2} )$ in the potential $U\left(
X\right) $~\cite{footnote}. It is remarkable that the soliton
propagates through the condensate without change of its density
profile. Indeed, simple calculation gives for the density
perturbation in the vicinity of the soliton:
\begin{equation}
\delta n(x) \equiv \left | \Psi(x) \right |^2-n =
-\frac{mu^2}{g_{1}} \frac{1}{\cosh^2 \left[\frac{1}{\ell}\left
(x-X \right ) \right]} \;.\label{profile}
\end{equation}
This quantity does not depend of time (at a given $x-X$), because
$u$ is an integral of motion.

Let us apply these results to a {\em harmonic} trap $U\left(
x\right) =m\omega_{x}^{2}x^{2}/2$. Notice that $\omega _{x}$ is
the frequency of oscillations of the center of mass of the
condensate.
Now the distribution of
the sound velocity is
$
c^{2}\left( X\right) =c_{0}^{2}\left( 1-X^{2}/L^{2}\right)
$
where $2L$ is the condensate length  defined by
$L^{2}=2\mu/m\omega_{x}^{2}$ and (\ref{final}) takes the form
$ \left( \frac{dX}{dt}\right) ^{2}=\frac{\mu }{m}-u^{2}
-\frac{\omega _{x}^{2}}{2}X^{2}\;. \label{fin}
$
This equation describes pure harmonic oscillations with frequency
$\omega _{s}={\omega _{x}}/{\sqrt{2}}$. This frequency
coincides with the one obtained in Refs.~\cite{Anglin,BK1} in a
different way for a slow soliton with $v \ll c_0$.
The amplitude of oscillations is equal to the coordinate $X_1$ of
the turning point:
\begin{equation}
X_{1}^{2}=\frac{2}{\omega _{x}^{2}}\left( \frac{\mu }{m}-
u^{2}\right ) =L^{2}-\frac{2}{\omega _{x}^{2}}u^{2}\;.
\end{equation}
The soliton oscillates between points $\pm X_1$, where the soliton
velocity and central density become zero. Our approach requires,
in addition to the general condition $\left( \ref{size}\right) $,
that the distance between  a turning point and the condensate
boundary is larger than the soliton width: $L-X_{1}\gg \ell $.
This condition can be transformed to
$
{u}/{c_{0}} \gg \left(
{\xi}/{L}\right ) ^{1/3}\;,\label{cond1}
$
which indicates that velocity of the soliton must not be too close
to speed of sound.

To conclude this section, we must make an important remark.
Equation (\ref{Es}) was derived for a uniform background in the
absence of an external potential. It is not obvious that the
energy has the same form in the presence of trapping potential.
The point is that a soliton is a region of decreased density and,
thus contains a negative "number of atoms"
\begin{eqnarray}
    \label{Ns}
    N_s[\Psi]=\int_{-\infty}^{\infty}\delta n(x)dx
    =-\frac{2\hbar
    }{g_1}u \;.
\end{eqnarray}
One may think that to the energy ${\cal E}$ one must add the
potential energy $N_sU(X) $ that corresponds to these atoms. Such
conclusion would be wrong. Actually, this decrease of energy is
compensated by increasing the energy of atoms outside of the
soliton. We will prove this in the next paragraph by direct
calculation of the energy.

It is worthwhile  to notice that the depletion (\ref{Ns}) of the
number of atoms in the soliton depends only on its energy ${\cal
E}$. This implies that $N_s$ is also an integral of motion. A
soliton is a quasiparticle of constant (negative) mass $mN_s$.
However, this "physical" mass of the soliton depends on energy and
is not equivalent to the "effective mass" $2m$.

\paragraph{Energy of a soliton in an external field.}

 In this paragraph we will show that equation (\ref{Es}) for the
energy of a soliton, obtained for an uniform condensate, is
actually valid also for a trapped condensate in the local density
approximation. Thus the trapping potential does not enter
explicitly in the expression for the energy of a soliton.

To take properly into account the expelling of atoms from the
soliton, it is more convenient to work not at constant number of
atoms, but at constant chemical potential $\mu $. Thus, instead of
the energy $E$, we will consider the "grand canonical energy"
$E^\prime = E-\mu\int n(x)dx$. In the GP approximation this energy
is $E^\prime=\int e^\prime dx$, where
\begin{equation}
e^\prime (x)=\frac{\hbar^2}{2m}|\Psi_x|^2
+\frac{1}{2}g_1n(x)^2+(U(x)-\mu)n(x)\;. \label{ex}
\end{equation}
Let the   soliton center be at $x=X$. We separate the integration
on two domains
$E^\prime=\int_{|x-X|>\delta} e^\prime dx +\int_{|x-X|<\delta}
e^\prime dx$
where $L\gg \delta \gg \ell$. In the first integral one can use
the TF approximation for the energy, neglecting the gradient term:
$e^\prime_{TF}(x)=\frac{1}{2}g_1n_{TF}(x)^2+(U(x)-\mu)n_{TF}(x)$,
while,  taking into account the smooth behavior of the potential,
in the second term one can approximate $U(x)\approx U(X)$. Next,
one can add to the first term an integral
$\int_{|x-X|<\delta} e_{TF}^\prime dx $ and, correspondingly,
deduct it from the second term in $E^{\prime}$. Such an addition
transforms the first term in the total energy of the condensate in
the absence of the soliton, which will be designated
$E^\prime_0$. In the second term one can safely change the smooth
function $e^\prime_{TF}(x)$ to $e^\prime_{TF}(X)$. Finally we
arrive at the result
\begin{widetext}
\begin{eqnarray}
E^\prime=E^\prime_0+\int_{|x-X|<\delta} \left [\frac{\hbar^2}{2m}|
\Psi_x |^2
+\frac{1}{2}g_1(n(x)^2-n_{TF}(X)^2)+(U(X)-\mu)(n(x)-n_{TF}(X))
\right ] dx\;.
 \label{Eprime0}
\end{eqnarray}
\end{widetext}

Now we can use the theorem about small increments (see~\cite{LL5},
\S 16). According to this theorem small corrections to $E$ and
$E^\prime $, are equal, being expressed, correspondingly, in terms
of $N$ and $\mu $. Thus to obtain a correction to the energy, we
must express the second term in (\ref{Eprime0}) in terms of the
density. One may eliminate $\mu $ from (\ref{Eprime0}) by taking
into account that, according to (\ref{nTF}),
$(\mu-U(X))=g_1n_{TF}(X)$. Owing to the fast convergence  of the
integral, it is also possible to change formally the integration
limits from $X \pm \delta $ to  to $\pm \infty $ respectivily.
Finally we get $E=E_0+{\cal E}$ where
\begin{equation}
{\cal E}=\int_{-\infty}^{\infty} \left [\frac{\hbar^2}{2m}|
\Psi_x |^2 +\frac{1}{2}g_1(n(x)-n_{TF}(X))^2 \right ] dx\;.
 \label{EpsilonF}
\end{equation}
Equation (\ref{EpsilonF}) does not contain the trapping potential
explicitly. Substituting  $\Psi(x) $  from (\ref{soliton}) and
integrating, we obtain equation (\ref{Es}), where $n$ is changed
to $n_{TF}(X)$. Equation (\ref{GP}) ensures now conservation
of {\cal E}.



\paragraph{Perturbative approach to grey-soliton dynamics.}

In the previous paragraphs investigating the soliton dynamics we
postulated, in the spirit of the Landau theory, that when the
soliton moves in a weakly-inhomogeneous background its local
energy stays constant. Let us now investigate the relation between
the above phenomenological approach and the evolution of a grey
soliton emerging from the dynamical approach, based on the
mean-field GP equation (\ref{GP}).

A grey soliton propagates against a background, and thus the first
step to obtain  the soliton dynamics is to determine the
background (which above was approximated by the TF density
distribution). One can do this by means of the ansatz~\cite{BK1}
$\Psi=\exp(-i\frac{\mu }{\hbar}t)F(x)\psi(x,t)$ where $F(x)$ is a
real-value solution of the nonlinear eigenvalue problem
\begin{eqnarray}
\label{backgorund} &&\mu
F=-\frac{\hbar^2}{2m}F_{xx}+U(x)F+g_1nF^3
\end{eqnarray}
with $F(0)=1,F_x(0)=0$ and $\lim_{x\to\pm\infty}F(x)=0$. As above
we stipulate that $U(0)=0$ and without loss of generality impose
the normalization condition on $F(x)$ at $x=0$ assuming that the
soliton is placed in the interval $x\in(-X_1,X_1)$, where $X_1 \ll
L$. The function $\psi(x,t)$ then solves the equation
\begin{eqnarray}
    \label{GP1}
    i\hbar\psi_t
    &+&\frac{\hbar^2}{2m}\psi_{xx}-g_1(|\psi|^2-n)\psi =R[\psi,F]
\end{eqnarray}
which is subject to the boundary conditions
$\lim_{x\to\pm\infty}|\psi(x,t)|=\sqrt{n}$. $R[\psi,F]$ is given
by
\begin{eqnarray}
    \label{R}
    R[\psi,F]\equiv-\frac{\hbar^2}{2m}\frac{F_x}{F}\psi_x+g_1n(F^2
    -1)(|\psi|^2-n)\psi\;.
\end{eqnarray}

An explicit form of Eq.(\ref{GP1}) appears to be useful if
$R[\psi,F]$ is small compared with the left hand side of that
equation. In order to estimate it we recall condition
(\ref{size}), which can be viewed as a definition of the small
parameter of the problem $\epsilon=\ell/L$. We restrict our
discussion to a soliton near the center of the trap, where
${|U(x)|}/{g_1n}={\cal O}(\epsilon)$ and the potential is smooth
enough, which enables one to neglect third order in the expansion
around $x=0$.  One can ensure then that $\mu\approx g_1n$ and
$R[\psi,F] ={\cal O}(\epsilon g_1n)$ in the domain specified
above.

This naturally leads one to use perturbation theory, the first
step of which would be an adiabatic approximation, i.e. an
approximation in which the  soliton shape is given by
(\ref{soliton}) with slowly varying parameters. This
straightforward approach, however meets difficulties in the case
at hand because of the first term in the right hand side of Eq.
(\ref{R}). Mathematical reasons for this are discussed
in~\cite{KV94,BK1}. Here we will use a more physical approach.
Namely, we will prove that it is possible to define an adiabatic
approximation for a dark soliton, such that in the leading order
the number of particles associated to the soliton is constant.
In other words, one has to prove that there exist a function
$\phi(x,t)$, satisfying (\ref{GP}) in the leading order and
preserving the quantity $N_s[\phi]=\int|\phi|^2dx$.
Because both $N_s$ and ${\cal E} $ are expressed in terms of the
same quantity $u$, the energy ${\cal E} $ is also an integral of
motion in accordance with the Landau theory. By direct algebra,
one may verify that the first term in the r.h.s. of (\ref{R})
results in change of the number of particles in the case of a
nonzero current [$\propto\int
\frac{F_x}{F}(\psi_x{\psi}^*-\psi{\psi}^*_x)dx$]. In order to
apply our arguments, we again assume that in the adiabatic
approximation the soliton is given by (\ref{soliton}). One can
then ensure that the imaginary part of all terms in (\ref{GP1})
becomes orecisely zero in the vicinity of the point $x=X(t)$
except for the first term in the r.h.s., which even grows with
$X(t)$. This behavior originates from non-adiabatic effects, which
must be taken into account before one passes to the Landau
quasiparticle description the soliton.

In order to avoid this difficulty, let us take into account that
the solution we are interested in is a function of $t$ and
$\zeta=x-X(t)$, where $X(t)$ is the coordinate of the soliton
center and can be associated with a coordinate in the Lagrangian
description of the condensate flow. Then, we make  the
substitution
\begin{eqnarray}
    \label{int_mode}
     \psi(x,t)=\phi(x,t)-if(t)\phi_x(x,t)\;,
\end{eqnarray}
where
\bigskip
\begin{eqnarray}
     f(t)=\frac{\hbar}{m}
     \int_{0}^{t}\frac{F_x(X(t'))}{F(X(t'))}\,dt'\;.
\end{eqnarray}
Here $\phi$ can be associated with the soliton wave function while
the second term on the r.h.s. of Eq. (\ref{int_mode}) is an
internal mode excited when a soliton moves in a potential.
Neglecting the terms of orders higher than $\epsilon $ one obtains
the equation  for $\phi(x,t)$
\begin{widetext}
\begin{eqnarray}
\label{GP2}
i\hbar\phi_t+\frac{\hbar^2}{2m}\phi_{xx}-g_1(|\phi|^2-n)\phi
 =-\frac{\hbar^2}{2m}\left[\frac{F_x(x)}{F(x)}-\frac{F_x(X(t'))}{F(X(t'))}\right]\phi_x
 +g_1n(F^2-1)(|\phi|^2-n)\phi-2if(t)g_1\phi^2{\phi}^*_x \;.
\end{eqnarray}
\end{widetext}
Now one can verify that $dN_s[\phi]/dt={\cal O}(\epsilon^2)$,
i.e  $\phi(x,t)$  describes a soliton with a constant mass.
Indeed, while the last two terms in the r.h.s. of (\ref{GP2})
obviously give zero contribution to the change of the number of
particles,
in order to treat the first one we recall the above requirement of
the smoothness of the potential $U(x)$. One then easily estimates
\[
\frac{F_x(x)}{F(x)}-\frac{F_x(X(t'))}{F(X(t'))}=\frac{(x-X)U_{xx}(0)}{2gn}+{\cal
O}(\epsilon^2).
\]
It is significant that the r.h.s. is an odd function of $x-X$. On
the other hand ${\phi}^*\phi_x-{\phi}^*_x\phi$, with $\phi $ given
by the r.h.s. of (\ref{soliton}), is an even function.
Thus, in computing $dN_s[\phi]/dt$, the first term contains an odd
function of $x-X$, which yields  zero to the leading order.

According to the previous reasoning,  the result obtained  proves
the conservation of the soliton energy.

\paragraph{Vortex ring.}
The method developed in this letter can be applied to other
localized excitations. The most interesting example is a vortex
ring. Let a small ring moves along the axis of the trapped
condensate. The energy of the ring in an uniform condensate can be
written as $\cal{E}=$ $\sqrt{n}\chi(v/c)$, where function $\chi$
is a decreasing function of its argument. (It was calculated in
\cite{Rob}, see also \cite{Pit1}, \S~5.4.) Then it follows from
conservation of $\cal{E} $ that when the ring moves in the
direction of decreased density, the ratio $v/c$ (and $v$ itself)
decreases and its radius increases. On the contrary, moving in the
direction of increasing $n$, the ring accelerates. If the central
density is big enough, the ring velocity can reach its maximal
value $v=0.93c$. The behavior  of the system near this point
cannot be described in the local density approximation. Probably
the excitation collapses beyond this point.

In conclusion, we have shown that considering a grey soliton as a
quasiparticle in the spirit of the Landau theory of superfluidity,
one can obtain a simple solution of the problem of soliton motion
in a trapped 1D condensate. The energy and the shape of the
soliton are preserved during its motion and soliton moves in a
trapping potential as a particle of mass $2m$.

\smallskip
We thank R. Benedek for helpful remarks. L.P. acknowledges support
from Argonne National Laboratory under contract with US DOE No.
W-31-109-ENG-38.


\begin{references}
\bibitem{footnote} Taking into account that energy ${\cal E }$
decreases when $v$ increases, it is more correctly to say that the
soliton moves as a particle of mass $(-2m)$ in the potential
$[-U(x)]$.
\end{references}
\end{document}